\documentclass[11pt]{article}
\def\be{\begin{equation}}
\def\ee{\end{equation}}
\def\ba{\begin{eqnarray}}
\def\ea{\end{eqnarray}}
\def\la{\langle}
\def\ra{\rangle}
\def\a{\alpha}
\def\b{\beta}

\def\h{\hskip 1cm}

\def\lo{\longrightarrow}
\def\A1{A_{-1}}
\usepackage{epsfig}
\begin{document}
\begin{titlepage}
\vspace{4cm}
\begin{center}{\Large \bf The matrix product representations for all valence bond states}\\
\vspace{2cm}\h Vahid Karimipour \footnote{Corresponding
author:vahid@sharif.edu},\h
Laleh Memarzadeh\footnote{email:laleh@physics.sharif.edu}, \\
\vspace{1cm} Department of Physics, Sharif University of Technology,
\\P.O. Box 11155-9161, Tehran, Iran
\end{center}
\vskip 2cm


\begin{abstract}
We introduce a simple representation for irreducible spherical
tensor operators of the rotation group of arbitrary integer or half
integer rank and use these tensor operators to construct matrix
product states corresponding to all the variety of valence-bond
states proposed in the Affleck-Kennedy-Lieb-Tasaki (AKLT)
construction. These include the fully dimerized states of arbitrary
spins, with uniform or alternating patterns of spins, which are
ground states of Hamiltonians with nearest and next-nearest neighbor
interactions, and the partially dimerized or AKLT/VBS (Valence Bond
Solid) states, which are constructed from them by projection. The
latter states are translation-invariant ground states of
Hamiltonians with nearest-neighbor interactions.
\end{abstract}

\hspace{.3in}
\end{titlepage}
\section{Introduction}\label{intro}
The problem of introducing exactly solvable models in quantum spin
chains has a long history in statistical mechanics and mathematical
physics, which encompasses a variety of models and techniques
\cite{Baxter,LiebMattis}. One can mention the $XY$ \cite{Mattis},
the Heisenberg $XXX$ and $XXZ$ \cite{Bethe}, the AKLT \cite{AKLT},
and the Majumdar-Ghosh models \cite{Majumdar}, and the free fermion,
the Bethe ansatz, and the matrix product techniques to name only a
few of the most important models and techniques which have been
developed so far. Finding any new exactly solvable model, is an
important step, since it acts as a reference model for developing
approximate perturbative solutions for more realistic models. It
will also help us to test many of the new ideas about collective
behavior of quantum systems, i.e. entanglement properties
\cite{korepin1, akm}, or the
relation of criticality and universality of entanglement \cite{Osterloh,NielsenOsborne}.\\

\begin{figure}[t]
 \centering
   \includegraphics[width=10cm,height=3.9cm,angle=0]{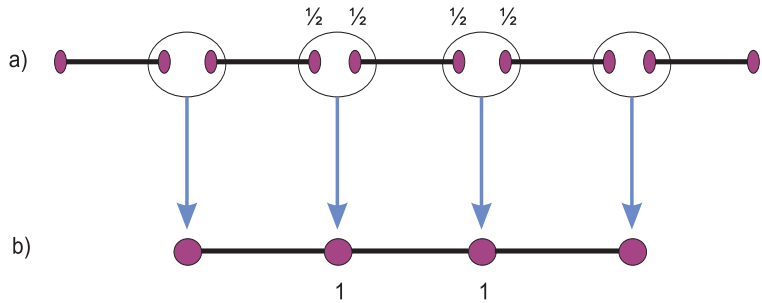}
   \caption{Color Online. A fully dimerized valence bond state, and the AKLT construction of
    a partially dimerized spin-1 state.
     The states in each bulb are projected to the symmetrized
     triplet. The parent Hamiltonian of the upper chain (a) has an
     interaction range of 3 lattice sites, while that of the lower
     chain (b) has a range of 2 sites.
    }
   \label{AKLToriginal}
\end{figure}

In \cite{AKLT}, Affleck, Kennedy, Lieb and Tasaki (AKLT), suggested
a new construction for a variety of spin states, known as valence
bond states. The basic element of this construction is a spin-1/2
singlet state, a dimer, $|s\ra=\frac{1}{\sqrt{2}}(|+,-\ra-|-,+\ra)$
which is called a valence bond in \cite{AKLT}. A dimerized state is
just a juxtaposition of such dimers on adjacent sites, figure
(\ref{AKLToriginal}-a). Such a state is clearly seen to be a ground
state of a Hamiltonian with three-sites interactions (nearest and
next-nearest neighbors), the local Hamiltonian of which is the
projector to spin 3/2 states, $h=P_{3/2}$. The reason is that due to
the presence of a dimer, the sum of spins of three adjacent sites
adds up only to spin 1/2. The parent Hamiltonian of this fully
dimerized state, is known as Majumdar-Ghosh Hamiltonian and has the
form
\begin{equation}\label{akltHamiltonian}
H=\sum_{j}\sigma_i\cdot \sigma_{i+1}+\frac{1}{2}\sigma_i\cdot
\sigma_{i+2}.
\end{equation}
This Hamiltonian has a two-fold ground state degeneracy, the other
ground state being simply a one-site translation of dimers to the
left or right.\\

\begin{figure}[t]
 \centering
   \includegraphics[width=8cm,height=4.1cm,angle=0]{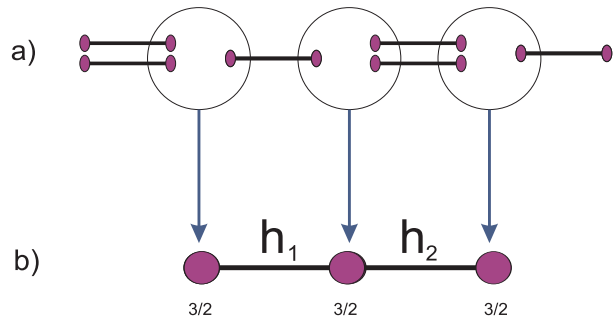}
   \caption{Color Online. An example of a dimerized state with alternating spin pattern and the AKLT construction of
   spin 3/2 chain. The interaction ranges are similar to that of figure (\ref{AKLToriginal}) .}
   \label{PartialDimersSimple}
\end{figure}

One can also consider fully dimerized states \cite{AKLT} with
alternating patterns of spins, where there are alternating number of
valence bonds or dimers. An example of this is shown in figure
(\ref{PartialDimersSimple}-a), where the local three-sites
Hamiltonian, should be taken as projector to spin 2, $h=P_2$.
Moreover one can use projection, to construct from these fully
dimerized states, partially dimerized or AKLT/VBS states which are
ground states of Hamiltonians with nearest-neighbor interactions.
For example in figure (\ref{AKLToriginal}-b), if one projects each
pair of spin-1/2 particles in a bulb of the original chain to the
symmetrized triplet, a non-dimerized spin-1 state is obtained on a
new chain, whose parent Hamiltonian which annihilates this state is
the sum of spin-2 projectors $P_2$ on consecutive sites. The reason
for this annihilation is that the sum of four initial spins on the
original chain (known also as the virtual chain) add up to at most
spin 1, due to the presence of the valence bond which is a singlet.
In this way a spin-1 quantum chain is obtained which is the exact
ground state of the following Hamiltonian:
\begin{equation}\label{akltHamiltonian}
H=\sum_{j}S_i\cdot S_{i+1}+\frac{1}{3}(S_i\cdot S_{i+1})^2.
\end{equation}

Projection can also be used for other types of dimerized state as
shown in figure (\ref{PartialDimersSimple}-b) to construct states
with arbitrary integer or half integer spins. For example in figure
(\ref{PartialDimersSimple}-b), looking at the number of valence
bonds which are singlets and are not counted in the addition of
spins in the virtual sites, one finds that the local Hamiltonians
can be chosen as $h_1=\lambda_3 P_3$ and $h_2=\lambda_3P_3+\lambda_2
P_2$, where $P_j$ is the projector on spin-$j$ states and
$\lambda_j$'s are positive coefficients. To assure translation
invariance for the parent Hamiltonian one then takes $H=\sum_j
h_{j,j+1}$ where $h=P_3$, is the operator common to both $h_1$ and
$h_2$. Needless to say, this construction can be generalized by
taking different alternating number of dimers in the virtual chain.
This is also the basic idea behind the exactly solvable spin-3/2
spin systems on the honeycomb lattice \cite{zit2d} or more generally
the basic idea behind PEPS, or Projected Entangled Pair States
\cite{peps},
which has only recently been discussed in the literature.\\

In the course of time, the basic idea of AKLT, which in turn was
inspired by the work of Majumdar and Ghosh \cite{Majumdar}, led to
the development of finitely correlated or matrix product
representation of states \cite{FCS1,FCS2,zitt1}, a representation
which when existing, greatly facilitates the calculation of many
properties of the ground states of quantum
systems \cite{akm,zitt1,MPS1,MPS2,MPS3,MPS4,Anders,AKS1,AKS3,mpsgeneral,zittartz32,Del,KolM}.\\
The Matrix Product (MP) representation was also found to be closely
related with the
success of density matrix renormalization group \cite{dmrg, dmrgcirac, McGulloch}.\\

When considering spin chains, the basic continuous symmetry is the
rotation symmetry captured by the $su(2)$ group, and there has been
many different and equivalent implementations of this symmetry in
matrix product states \cite{AKS1,AKS3,Del,KolM,McGulloch}. While a
lot of progress has been made in defining matrix product states,
having specific symmetries, to our knowledge the original AKLT
variety of states, have not been cast into a simple and uniform
matrix product form for both integer and non-integer spins. For the
integer case however, such a formulation has been reported in
\cite{as}. There is no doubt that such a representation, will be of
utmost importance for further study of AKLT models, and even for
similar models on more general geometries like the
Bethe Lattice \cite{shor}.\\

To ensure invariance of the Matrix Product State (MPS) under
rotation, it is sufficient that the elementary matrices used in the
definition of the MPS constitute a representation of spherical
tensor operators of a specific rank. The rank of the tensor depends
on the spin of the actual lattice and the dimension of the
representation determines the dimension of the auxiliary matrices.
Finding a simple and minimal-dimensional representation for such
tensors, constitute the basic problem in constructing rotationally
invariant MPS, both for spin chain, spin ladders, or two dimensional
lattices.\\

What we will do in this paper is to provide a uniform and simple
matrix product representation for all the AKLT, or valence bond
states and even more general states. The starting point of our
analysis is a simple and compact representation of spherical tensor
operators of any rank, integer or half integer. These tensors enable
us to define MP representations for Majumdar-Ghosh states, (which
are the ancestors of AKLT states) and their generalization to
arbitrary spins, and then we will use them to construct MP
representation for partially dimerized states. We then use
projection method to find MP representations for arbitrary spin
chains, with nearest neighbor interaction. The parallel with the
AKLT construction is simple: the basic idea is to replace a
collection of $2s$ spin-1/2 dimers or valence bonds with a single
spin-$s$ valence bond and represent the states constructed from
these
spin-$s$ valence bonds as MPS.\\

Besides having the benefit of calculability, when we have an MPS
representation, the very method of MPS allows us to find a larger
family of Hamiltonians than the AKLT method. This larger family,
with its larger number of couplings will enable us to better adjust
or approximate an exactly solvable Hamiltonian with realistic
situations. We will see an example of this in this paper.\\

The structure of this paper is as follows: In section \ref{MPS} we
review the matrix product formalism \cite{FCS1,FCS2,zitt1} in a
language which we find convenient \cite{AKS1} for further
developments. In particular we emphasize the symmetry properties of
the ground state and the Hamiltonian. In section \ref{dimerModel} we
will introduce a compact formula for spherical tensors of rank $s$
(integer or half integer) and use it to construct dimerized states
of arbitrary integer or half-integer spins in section
(\ref{FullDimer}). These are the generalization of Majumdar-Ghosh
states, or fully dimerized states, to arbitrary spins. We then go on
in section (\ref{NNModel}) to define MP representations for other
types of dimerized states. In section (\ref{NONDimers}) we find MP
representations for AKLT/VBS states. The core of this section is the
definition of new kinds of tensors, which play the role of auxiliary
matrices for the MP representations of these states. Section
(\ref{Examples}) is devoted to some specific examples, where more
detailed properties of some of the states and their parent
Hamiltonians are derived. We conclude the paper with a discussion.

\section{Matrix Product States}\label{MPS}
Let us first make a quick review of the matrix product states in a
language which we find convenient \cite{AKS1,AKS3}. For more
detailed reviews of the subject, the reader can consult a more
comprehensive review article like \cite{mpsgeneral} or any of the
many works where specific examples have been studied
\cite{akm,AKS1,AKS3,MPS1,MPS2,MPS3,Anders,MPS4,zittartz32,Del,KolM}.

Consider a ring of $N$ sites, where each site describes a $d-$level
state. The Hilbert space $C^d$ of each site is spanned by the basis
vectors $|i\ra, \ \ i=0,\cdots , d-1$. A state
\begin{equation}\label{state}
    |\Psi\ra=\sum_{i_1,i_2,\cdots, i_N}\Psi_{i_1i_2\cdots
    i_N}|i_1i_2\cdots i_N\ra
\end{equation}
is called a matrix product state if there exist $D$ dimensional
matrices $A_i\in C^{D\times D},\ \ i=0,\cdots, d-1$, such that
\begin{equation}\label{mps}
    \Psi_{i_1i_2\cdots
    i_N}=\frac{1}{\sqrt{Z}}tr(A_{i_1}A_{i_2}\cdots A_{i_N}),
\end{equation}
where $Z$ is a normalization constant. This constant is given by $
Z=tr(E^N), $  where $ \label{E} E=\sum_{i=0}^{d-1} A_i^*\otimes
A_i$. Note that we are here considering homogeneous matrix product
states  where the matrices depend on the value of the spin at each
site and not on the site itself. More general MPS's can be defined
where the matrices depend also on the position of the sites
\cite{mpsgeneral}. \\

The collection of matrices $\{A_i\}$ and $\{\mu UA_iU^{-1}\}$, where
$\mu$ is an arbitrary complex number, both lead to the same matrix
product state, the freedom in scaling with $\mu$, is due to its
cancelation with $Z$ in the denominator of (\ref{mps}).  This
freedom will be useful when we
discuss symmetries. There has been discussions on the symmetry of matrix product states in the literature
\cite{FCS1,FCS2,MPS4,AKS1,Del,KolM}, here we use the language or notation used in \cite{AKS1}.\\

\subsection{Symmetries of the ground state}
Consider a local continuous symmetry operator $R$ acting on a site
as $R|i\ra=R_{ji}|j\ra$ where summation convention is being used.
$R$ is a $d$ dimensional unitary representation of the symmetry. A
global symmetry operator ${\cal R}:=R^{\otimes N}$ will then change
this state to another matrix product state
\begin{equation}\label{mpsPrime}
    \Psi_{i_1i_2\cdots i_N}\lo \Psi'_{i_1 i_2\cdots i_N}:=tr(A'_{i_1}A'_{i_2}\cdots
    A'_{i_N}),
\end{equation}
where
\begin{equation}\label{A'}
    A'_i:=R_{ij}A_j.
\end{equation}
The state $|\Psi\ra$ is invariant (i.e. a singlet) under this
symmetry if there exist an operator $U(R)$ such that
\begin{equation}\label{symm}
    R_{ij}A_j=U^{-1}(R)A_iU(R).
\end{equation}
Repeating this transformation and using the group multiplication of
the transformations $R$, puts the constraint
\begin{equation}\label{UR}
    U_{R'}U_R=U_{R'R}.
\end{equation}
Thus $U(R)$ is a $D$ dimensional representation of the symmetry $R$.
In case that $R$ is a continuous symmetry with generators $T_a$,
equation (\ref{symm}), leads to
\begin{equation}\label{symmalg}
    (T_a)_{ij} A_j=[{\cal T}_a , A_i],
\end{equation}
where $T_a$ and ${\cal T}_a$ are the $d-$ and $D-$dimensional
representations of the Lie algebra of the symmetry. \\

\subsection{Symmetries of the Hamiltonian:} Given a matrix product state, the
reduced density matrix of $k-$ sites is given by
\begin{equation}\label{rhoksites}
    \rho_{i_1\cdots i_k, j_1\cdots j_k}=\frac{tr((A_{i_1}^*\cdots A_{i_k}^*
    \otimes A_{j_1}\cdots A_{j_k})E^{N-k})}{tr(E^N)}.
\end{equation}
The null-space of this reduced density matrix contains the subspace
spanned by the solutions of
\begin{equation}\label{cAA}
    \sum_{j_1,\cdots, j_k=0}^{d-1}c_{j_1\cdots
    j_k}A_{j_1}\cdots A_{j_k}=0.
\end{equation}

Let the null space of the reduced density matrix of $k$ adjacent
sites, denoted by $\Delta_k$, be spanned by the orthogonal vectors
$|e_{\a}\ra, \ \ \ (\a=1, \cdots, s\geq d^k-D^2)$. Then we can
construct the local hamiltonian acting on $k$ consecutive sites as
\begin{equation}\label{h}
    h:=\sum_{\a=1}^s \lambda_{\a} |e_{\a}\ra\la e_{\a}|,
\end{equation}
where $\lambda_{\a}$'s are positive constants. These constants
together with the parameters of the vectors $|e_{\a}\ra $ inherited
from those of the original matrices $A_i$, determine the total
number of coupling constants of the Hamiltonian. If we call the
embedding of this local Hamiltonian into the sites $l$ to $l+k$ by
$h_{l,l+k}$ then the full Hamiltonian on the chain is written as
\begin{equation}\label{H}
    H=\sum_{l=1}^N h_{l,l+k}.
\end{equation}
The state $|\Psi\ra$ is then a ground state of this hamiltonian with
vanishing energy. See \cite{AKS1} for a more detailed discussion of
the above points.\\

A Hamiltonian derived as above does not have any particular
symmetry. Indeed the above class include all types of Hamiltonians
which have the matrix product state as their ground state.  A
subclass of these Hamiltonians however do have the symmetry of the
ground state. Consider equation (\ref{cAA}), multiplying both sides
of this equation by $U^{-1}(R)$ from left and $U(R)$ from right, and
using (\ref{symm}), we find that if $c_{i\cdots j}$ is a solution of
(\ref{cAA}), then $R_{i,i'}\cdots R_{j,j'}c_{i',\cdots j'}$ is also
a solution of the same equation, that is:
\begin{equation}\label{symmH}
    R_{i,i'}\cdots R_{j,j'} c_{i',\cdots j'}=\lambda(R) c_{i,\cdots
    j}.
\end{equation}
This means that the null space of the reduced density matrix is an
invariant subspace under the action of the symmetry group
$R^{\otimes k}$. Thus the null vectors $|e_{\a}\ra$ transform into
each other under the action of the reducible representation
$R^{\otimes k}$. Such vectors can be classified into multiplets such
that each multiplet transforms under one irreducible representation
of the group $R$. Let the states transforming under the irreducible
representation $D^{\mu}$ of the group, be denoted by
$|e^{\mu}_{\b}\ra$. Then the operators
$h^{\mu}=\sum_{\b}|e^{\mu}_{\b}\ra\la e^{\mu}_{\b}|$, is a scalar
under the action of the group, that is
\begin{equation}\label{Dmu}
    [D^{\mu}, h^{\mu}]=0.
\end{equation}
Hence to ensure the symmetry of the local Hamiltonian we write it as
\begin{equation}\label{Hsymm}
    h_{i,i+k}=\sum_{\mu}{\lambda_{\mu}}h^{\mu}=\sum_{\mu,\b}\lambda_{\mu}|e^{\mu}_{\b}\ra\la
    e^{\mu}_{\b}|,
\end{equation}
where the number of free couplings $\lambda_{\mu}$ is equal to the
number of multiplets which span the null space $\Delta_k$.

\section{A new representation for spherical tensors of arbitrary rank}\label{dimerModel}
We are now equipped with generalities about matrix product states
and their symmetry properties. In this section we specialize the
above discussion to construction of spin-s MPS invariant under
rotation in spin space. For such a chain we take local Hilbert space
to be spanned by the $d=2s+1$ states of a spin-$s$ particle, i.e.
the states $\{|s,m\ra, m=-s,\cdots s\}$. Let us denote the $D$
dimensional matrix assigned to the local configuration $m$ by
$A_{s,m}$. Rotational symmetry in the spin space now demands that
the matrices $A_{s,m}$ form an irreducible tensor operator of rank
$s$ in the space of $D$ dimensional square matrices. In view of
(\ref{symmalg}), we should find $2s+1$ matrices $A_{s,m}$ such that
the following relations are satisfied
\ba\label{RotSym}
[L_z,A_{s,m}]&=&mA_{s,m}\cr
[L_+,A_{s,m}]&=&\sqrt{s(s+1)-m(m+1)}A_{s,m+1}\cr
[L_-,A_{s,m}]&=&\sqrt{s(s+1)-m(m-1)}A_{s,m-1} \ea

where $L_z$, $L_+$ and $L_-$ are the $D$ dimensional representations
(not necessarily irreducible) of the Lie algebra of $su(2)$:

\begin{equation}
  [L_z,L_{\pm}] = \pm L_{\pm} \h
  [L_+,L_-] = 2L_z.
\end{equation}

\textbf{Remark:} For simplicity, we will use the notation $|m\ra$
and $A_m$ instead of $|s,m\ra$ and $A_{s,m}$ respectively, when the
label $s$ is clear from the context.\\

It is crucial to note that it is not always possible to find tensor
operators of a given rank for a given $D$ dimensional
representation. For example while there is tensor of rank one, in
two dimensions, given by

\begin{equation}\label{rank1tensor}
    A_{1}=-\sqrt{2} \sigma_+, \ \ \ \ A_0 = \sigma_z, \ \ \ \
    A_{-1}=\sqrt{2} \sigma_-,
\end{equation}
leading to the spin-1 AKLT model \cite{AKLT}, with parent
Hamiltonian (\ref{akltHamiltonian}), there is no rank $1/2$ tensor
operator in $2$ dimensions. By this we mean that if we take $D=2$,
then there is no non-zero solution for the following system of
matrix equations
\begin{eqnarray}\label{rankonehalftensor}
  [L_z,A_{\pm}] &=& \pm\frac{1}{2} A_{\pm} \cr
  [L_+,A_{+}] &=& 0,\ \ \ \ [L_+,A_-]=A_+,\cr
  [L_-,A_{+}] &=& A_-,\ \ \ \ [L_-,A_-]=0.
\end{eqnarray}

Therefore the first task for construction of rotationally invariant
matrix product states for quantum spin chains or quantum ladders is
to have a compact expression for spherical tensors of arbitrary
rank.

A possible procedure for obtaining spherical tensors of integer rank
is to take two low-rank (possibly identical) tensors and decompose
their ordinary or tensor product by the Clebsh-Gordon series to
obtain irreducible tensors of higher rank. In fact if $A_{s,m}$ and
$A_{s',m'}$ are two spherical tensors, then one can form the product
$A_{s,m}A_{s',m'}$ (if their dimensions are the same) or
$A_{s,m}\otimes A_{s',m'}$ (otherwise) and decompose the products by
using the Clebsh-Gordon coefficients to obtain spherical tensors of
higher rank. For example take the AKLT tensor of rank one. Ordinary
multiplication of this tensor, does not give a tensor of rank 2,
since $\sigma_+^2=0$, however its tensor multiplication gives a
tensor of rank two of dimension 4, i.e. $A_{2,2}=2\sigma_+\otimes
\sigma_+$, etc. In this way the product of two rank-1 tensors can be
decomposed to give a rank-2, a rank-1 and rank-0 tensor. The
obtained tensors can again be multiplied with other tensors and
decomposed to obtain tensors of even higher rank. This procedure
however has several drawbacks: first the dimensions of the matrices
will grow very fast as we increase the rank of tensors, second it
requires  multiple use of Clebsh-Gordon coefficients which makes the
final expression of the tensors, especially for high-rank tensors,
quite cumbersome and not useful. Another useful procedure, is to
invoke the Wigner-Eckart theorem which decomposes the matrix
elements of any spherical tensor in the angular momentum basis, to
an angular part, which is the Clebsh-Gordon coefficient and a
reduced part, which essentially defines the tensor. However this
procedure does not always lead to a compact notation for the tensor
operators themselves and the multiplication of such tensors requires
heavy use of Glebsh-Gordon coefficients. In this paper we introduce
a compact and transparent formula for spherical tensors of rank $s$,
for $s$ integer or half integer, and use it to construct matrix
product states for spin chains. For rank-$s$ tensors the dimensions
of the matrices are $2s+2$, thus the dimension grows linearly with
rank. \\

To construct the spherical rank-$s$ tensor, let us take the
orthonormal basis $\{|s,m\ra,\ m=-s\cdots s \}$ of the spin $s$
representation and augment it by the single state $|\tilde{0}\ra$,
of the spin $0$ representation
\begin{equation}\label{aux}
    \la \tilde{0}|\tilde{0}\ra=1, \ \ \ \ \la \tilde{0}|s,m\ra=\la
    s,m|\tilde{0}\ra=0.
\end{equation}
On this larger space, the following is the reducible $s\oplus 0$
representation of angular momentum algebra:
\begin{eqnarray}
&&L_z=\sum_{m=-s}^s m|s,m\ra\la s,m| \cr &&L_+ = \sum_{m=-s}^s
\sqrt{{s(s+1)-m(m+1)}}|s,m+1\ra \la s,m|\cr &&L_- = \sum_{m=-s}^s
\sqrt{{s(s+1)-m(m-1)}}|s,m-1\ra \la s,m|.\end{eqnarray}

Now it is readily verified that in this $2s+2$ dimensional space,
the following matrices form an irreducible rank-$s$ spherical
tensor, that is they satisfy the relations (\ref{RotSym}):
\begin{equation}\label{defA}
A_{s,m}:=|s,m\ra\la\ \tilde{0}|+(-1)^{s-m}|\tilde{0}\ra\la s,-m|
\end{equation}
where $-s\leq m \leq s$. \\
It is important to note that the rank of these tensors can be
integer or half integer. Such operators transform as an irreducible
rank $s$ tensor in the space which carries the reducible
representation $s\oplus 0$.\\

\textbf{Note:} One can define the tensors more generally as
$$A_{s,m}:=a|s,m\ra\la\ \tilde{0}|+b(-1)^m|\tilde{0}\ra\la s,-m|$$
where $a$ and $b$ are arbitrary numbers, however these tensors are
equivalent to the previous ones in the sense that they reduce to
them by a suitable unitary transformation. The factor $(-1)^s$ is
inserted in the definition to ensure that no complex number enters
the expression for half-integer ranks.\\

While there are many representations for spherical tensors of
different ranks, and these have been used in different works to
construct various examples of invariant MPS
\cite{MPS1,MPS2,MPS3,MPS4,AKS1,Del,KolM}, to our knowledge the
representation (\ref{defA}) is introduced for the first time. In the
sequel we show that this representation is very general, in the
sense that we can use it to find MP representations for all the
variety of AKLT states, including the Majumdar-Ghosh or fully
dimerized states of arbitrary spin, the partially dimerized states,
and also the various states which are found from these partially
dimerized states by different types of projection. Even more, one
can construct other states not listed in the original AKLT papers,
these are the symmetry breaking states.

\section{The spin-s fully dimerized or Majumdar-Ghosh
states}\label{FullDimer} Using the definition of $A_m$ we find:
\ba\label{ProVm} A_{m_1}A_{m_2}\cdots
A_{m_{2N}}&=&\prod_{i=1}^{N}(-1)^{s-m_{2i-1}}\delta_{m_{2i-1},-m_{2i}}
|\tilde{0}\ra\la\tilde{0}|\cr
&+&(-1)^{s-m_{2N}}\prod_{i=1}^{N-1}(-1)^{s-m_{2i}}\delta_{m_{2i},-m_{2i+1}}|m_1\ra\la-m_{2N}|
\ea Taking the trace we find
\begin{eqnarray}\label{trace}
Tr(A_{m_1}\cdots
A_{m_{2N}})=\prod_{i=1}^{N}(-1)^{s-m_{2i-1}}\delta_{m_{2i-1},-m_{2i}}
&+& \prod_{i=1}^{N}(-1)^{s-m_{2i}}\delta_{m_{2i},-m_{2i+1}}.
  \end{eqnarray}
Inserting this into (\ref{state}-\ref{mps}) the final simple form of
the matrix product ground state is obtained as
\begin{figure}[t]
 \centering
   \includegraphics[width=9cm,height=3cm,angle=0]{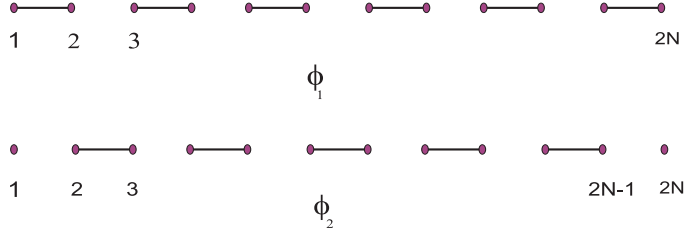}
   \caption{Color Online. The fully dimerized spin-s states in equation (\ref{gmps}).
   }
   \label{new3}
\end{figure}

\begin{equation}\label{gmps}
|\psi\ra\equiv |\phi_1\ra+|\phi_2\ra=|S\ra_{12}|S\ra_{34}\cdots
|S\ra_{2N-1,2N}+|S\ra_{23}|S\ra_{45}\cdots|S\ra_{2N,1},
\end{equation} where the singlet states $|S\ra$ are given by
\begin{equation}\label{SSS}
    |S\ra=\frac{1}{\sqrt{2s+1}}\sum_{m}(-1)^{s+m} |m,-m\ra.
\end{equation}

Note that $|S\ra$ is a singlet state, i.e. $L_z|S\ra =
L_+|S\ra=L_-|S\ra=0$. Thus $|\phi_1\ra$ is a juxtaposition of spin-s
dimers on sites $(1,2), (3,4), \cdots (2N-1,2N)$ and $|\phi_2\ra$ is
a one-site translation of $|\phi_1\ra,$ i.e. a collection of spin-s
dimers on sites $(2,3), (4,5),\cdots (2N,1)$, figure (\ref{new3}).

\section{Other types of dimerized states}
\label{NNModel}

\begin{figure}[t]
 \centering
   \includegraphics[width=12cm,height=3.6cm,angle=0]{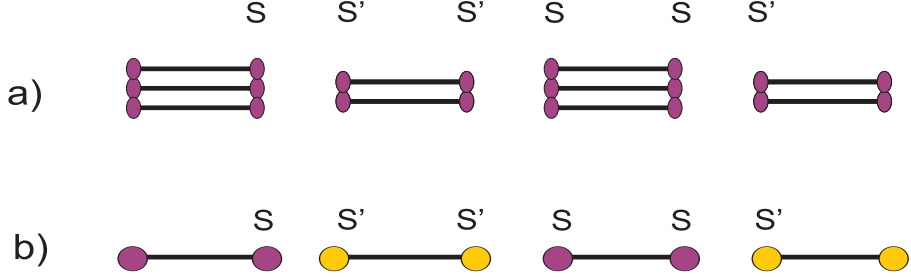}
   \caption{Color Online.
   2$s$ valence bonds (spin 1/2 singlets) in the AKLT construction,
   are replaced in our work, with a single spin-s valence bond,
   which is MPS representable by the matrix
   (\ref{pa}).}
   \label{PartialDimersCompared}
\end{figure}

A general dimerized state is one which is shown in figure
(\ref{PartialDimersCompared}-a), where each line stands for a
spin-1/2 dimer. The numbers of dimers are $2s$ and $2s'$
respectively. In our representation, we replace 2s spin-1/2 dimers
with a single spin-s dimer, as in figure
(\ref{PartialDimersCompared}-b).
 Such a state has simple MPS representation, in the form
\begin{equation}\label{pa}
|\psi\ra=\sum
tr(A_{m_1}A_{m_2}B_{m'_1}B_{m'_2}A_{m_3}A_{m_4}B_{m'_3}B_{m'_4}\cdots)|m_1
m_2 m'_1 m'_2 m_3 m_4 m'_3 m'_4\cdots\ra\ ,
\end{equation}
where the matrices $\{A_m\}$ and $\{B_{m'}\}$ are embedding of the
rank-$s$ and rank-$s\prime$ tensors (\ref{defA}) into a
representation spanned by the vectors
$\{|\tilde{0}\ra,|s,m\ra,|s',m'\ra\}$, i.e. the direct sum
representation $s\oplus s'\oplus 0$. In fact it is readily found
that with
\begin{eqnarray}
A_{s,m}\equiv A_m&=&|s,m\ra\la\tilde{0}|+(-1)^{s-m}|\tilde{0}\ra\la
s,-m|,\cr A_{s',m'}\equiv
B_{m'}&=&|s',m'\ra\la\tilde{0}|+(-1)^{s'-m'}|\tilde{0}\ra\la
s',-m'|,
\end{eqnarray}
we have \be
A_{m_1}A_{m_2}B_{m'_1}B_{m'_2}=(-1)^{s-m_1}\delta_{m_1,-m_2}(-1)^{s'-m'_1}\delta_{m'_1,-m'_2}
|\tilde{0}\ra\la\tilde{0}| \ee which readily yields the following
partially dimerized form for the state (\ref{pa}):
\begin{eqnarray}
|\psi\ra=|S\ra_{12}|S'\ra_{34}|S\ra_{56}|S'\ra_{78}\cdots
\end{eqnarray}
where $ |S\ra $ and $|S'\ra$ are respectively spin-$s$ and spin-$s'$
singlets defined in (\ref{SSS}).\\

\begin{figure}[t]
 \centering
   \includegraphics[width=8cm,height=1cm,angle=0]{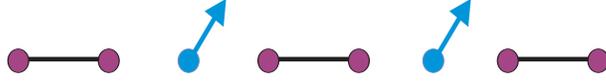}
   \caption{Color Online. The tensors (\ref{defA}) can also be used to construct symmetry-breaking matrix product states.
   .}
   \label{SymmetryBreakingFigure}
\end{figure}

One can construct even more general states, i.e. the symmetry
breaking states of the form shown in figure
(\ref{SymmetryBreakingFigure}) where the dimers are interspaced by
spins which align in a particular direction. Consider the state
\begin{equation}\label{SymmetryBreaking}
    |\psi\ra=tr(A_{_{m_1}}A_{_{m_2}}C_{_{m_3}}A_{_{m_4}}A_{_{m_5}}C_{_{m_6}}\cdots)|m_1,m_2,m_3,m_4,m_5,m_6,\cdots\ra
\end{equation}

where $A_m$ is of the form (\ref{defA}) and
$C_m=\alpha_m|\tilde{0}\ra\la \tilde{0}|$, in which $\a_m$'s
($-s\leq m\leq s$) are arbitrary complex numbers. Then the MPS
represents a symmetry breaking state shown in figure
(\ref{SymmetryBreakingFigure}), where spins, $3, 6, \cdots$ are
aligned in the state $|\a\ra:=\sum_{m=-s}^s\a_m|s,m\ra$ and the rest
of the sites are dimerized.  A suitable projection of these states,
gives symmetry-breaking non-dimerized states \cite{abk}.

\section{The AKLT/VBS states}\label{NONDimers}
\begin{figure}[t]
 \centering
   \includegraphics[width=10cm,height=5cm,angle=0]{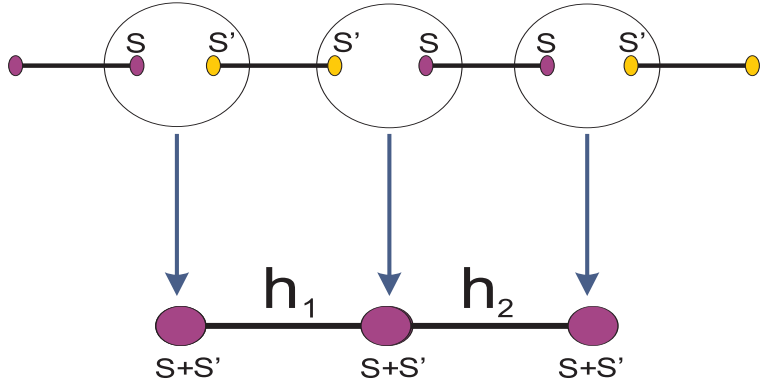}
   \caption{Color Online. The projection method: From a partially or fully dimerized
   state, whose parent Hamiltonian has interaction range 3,
    an AKLT/VBS state is constructed whose parent Hamiltonian has interaction range 2.
    The form of the Hamiltonian is given in equations (\ref{Haklt}) and (\ref{haklt}).}
   \label{OurPartialDimers}
\end{figure}

In the AKLT models, one can use the fully dimerized states and
project them to states which are called VBS states. While the parent
Hamiltonian of the fully dimerized states has an interaction range
of 3 sites, the VBS states which are obtained by projection have
parent Hamiltonians with interaction range of 2 sites. The method is
explained in figure (\ref{OurPartialDimers}), where we use a single
spin-s dimer to replace 2s spin-1/2 dimers in the original method of
AKLT.

The lower state is obtained by projecting the states inside each
bulb in the upper chain onto the symmetrized spin sector with total
spin $s+s'$. It is now obvious how the parent Hamiltonian of the
lower chain, the Hamiltonian which has this state as its ground
state, should be constructed. Consider the first bond in figure
(\ref{OurPartialDimers}) whose local Hamiltonian is denoted by
$h_1$. Due to the $2s'$ singlets, between the two bulbs, here we are
only summing over two spin $s$ states, instead of the apparent two
spin $s$ and two spin-$s'$ states. Hence all the projectors $P_j$,
with $2s+1\leq j\leq 2s+2s'$, annihilate this bond, i.e. the local
Hamiltonian $h_1$, can be constructed as a linear superposition of
all the above projectors with positive coefficients. By the same
reasoning the local Hamiltonian $h_2$ can be a linear superposition
of all projectors $P_k$, with $2s'+1\leq k\leq 2s+2s'$. Thus to
construct a translation-invariant Hamiltonian, the parent
Hamiltonian of the lower state can be constructed as
\begin{equation}\label{Haklt}
    H=\sum_i h_{i,i+1},
\end{equation}
where
\begin{equation}\label{haklt}
    h=\sum_{j=max(2s,2s')+1}^{2s+2s'}\lambda_j P_j,
\end{equation}
where $P_j$'s are the projectors on spin $j$ sector of two sites and
$\lambda_j$ are positive coefficients.

Note that the state on the lower chain is no longer dimerized, i.e.
spins which are further apart than one lattice spacing, are
correlated. Needless to say, the projection method, although elegant
in principle, is not suitable for calculation of many properties of
the state. Having a matrix product representation for this state,
turns all calculations into a straightforward and handy procedure.
In this section we show that the irreducible tensors introduced in
section (\ref{dimerModel}), provides a MP representation for these
states in a very simple way.

The starting point of our procedure is however not to use $2s$ and
$2s'$ spin-1/2 dimers as in figure (\ref{PartialDimersCompared}-a),
rather we use equivalently one spin-s and one spin-s' singlets as in
figure (\ref{PartialDimersCompared}-b), for which we have already a
MP representation. The spin-s and spin-s' dimers come from rank-s
and rank-s' tensors (\ref{defA}). We multiply and symmetrize these
two tensors to obtain a new tensor whose highest component is given
by
\begin{equation}\label{AB}
    V_{s+s',s+s'}=A_{s,s}A_{s',s'}+A_{s',s'}A_{s,s}.
\end{equation}

From the explicit form of the tensors in (\ref{defA}), one sees
that,
\begin{equation}\label{Vss}
    V_{s+s',s+s'}:=|s,s\ra\la s',-s'|+|s',s'\ra\la s,-s|.
\end{equation}
Note that this tensor lives in the $2s+2s'+2$ dimensional space
spanned by independent vectors $\{|s,m\ra,\ |s',m'\ra,\ -s\leq m\leq
s,\ -s'\leq m'\leq s'\}$.

It is readily verified that
\begin{eqnarray}\label{LVss}
    [L_z,V_{s+s',s+s'}]&=&(s+s')V_{s+s',s+s'}\cr
    [L_+,V_{s+s',s+s'}]&=&0.
\end{eqnarray}
Therefore $V_{s+s',s+s'}$ is indeed the highest-weight component of
a spherical tensor of rank $s+s'$. Other components are obtained by
successive commutations with $L_-$. For example, we have
\begin{eqnarray}\label{Vss2}
    V_{s+s',s+s'-1}&:=&\sqrt{\frac{s}{s+s'}}(|s,s-1\ra\la s',-s'|-|s',s'\ra\la s,-s+1|)\cr&+&\sqrt{\frac{s'}{s+s'}}
    (|s',s'-1\ra\la s,-s|-|s,s\ra\la s',-s'+1|).
\end{eqnarray}
The new spherical tensors have the interesting property that they
lead to a non-empty null space $\Delta_2$. In fact it can be
verified that these tensors have a peculiar fusion rule
(decomposition of the product into irreducible representations),
which exactly matches the fusion rule of the original $2s$ and $2s'$
valence bonds in a symmetric way. In the present formalism, this
symmetry causes the final local Hamiltonian $h$ to contain
projectors common to both $h_1$ and $h_2$ (figure
(\ref{OurPartialDimers})) in the AKLT construction. Using the
notation $V_s$ to denote the whole multiplet $V_{s,m}, \ -s\leq
m\leq s$, the fusion rule
 of
our tensors is
\begin{equation}\label{fusion}
    V_{s+s'}\otimes V_{s+s'}=\bigoplus_{j=0}^{max(2s,2s')} c_j V_{j}.
\end{equation}
Thus the multiplets $V_j$ with $max(2s,2s')+1\leq j\leq 2(s+s')$ are
absent, i.e. identically vanish, in the decomposition of the left
hand side tensors. In the language of matrix product formalism,
section (\ref{MPS}), this means that the null-space of the two-site
density matrix, contains the multiplet of states which transform as
spin $j$ representations with $max(2s,2s')+1\leq j\leq 2(s+s')$.
Therefore the local Hamiltonian annihilating the dimerized state,
can be constructed from the projectors to these multiplets, namely
\begin{equation}\label{hh}
    h=\sum_{j=max(2s,2s')+1}^{2s+2s'}\lambda_j P_j,
\end{equation}
where $P_j$'s are projectors on spin $j$ and $\lambda_j$ are
positive coefficients. It requires tedious and lengthy calculations
which may not be illuminating to prove (\ref{fusion}) in general.
Instead we will give an idea of the proof by way of examples. First
of all, it is readily seen from (\ref{Vss}) that $s,s'\ne 0,$
$$V_{s+s',s+s'}^2=0,$$ but this is the top state of the multiplet $V_{2s+2s'}$  and hence this multiplet
is absent in the right hand side of (\ref{fusion}). In the same way
one can also show from (\ref{Vss}) and (\ref{Vss2}) that the top
state of the multiplet $V_{2s+2s'-1}$ is zero. This pattern repeats
until we arrive at the multiplet $V_{max(2s,2s')}$. We will give a
more detailed and concrete example in section (\ref{Examples}).

\section{Examples}\label{Examples}
Up until now we have been able to use our spherical tensors
(\ref{defA}), in a uniform manner to construct all the variety of
valence-bond states in the AKLT constrution.  In this section, we
will provide a few concrete examples.

\subsection{Properties of fully dimerized or spin-s Majumdar-Ghosh
states} First we calculate the normalization of fully dimerized
states $|\Psi^{\pm}\ra.$ The basic tool which we use is the
following easily verified equation between the singlets, where
$1,2,3$ and $4$ are any four different and not necessarily adjacent
sites:
\begin{equation}\label{SingletProduct}
    _{_{23}}\la S|S\ra_{_{12}}|S\ra_{_{34}}=\frac{\epsilon}{2s+1}|S\ra_{_{14}},
\end{equation}
where $$\epsilon= \left\lbrace\begin{array}{ll} 1 & s={\rm
integer}\\
\\ -1 & s={\rm half \ integer}\end{array}\right.
$$

\begin{figure}[t]
 \centering
   \includegraphics[width=8cm,height=1.25cm,angle=0]{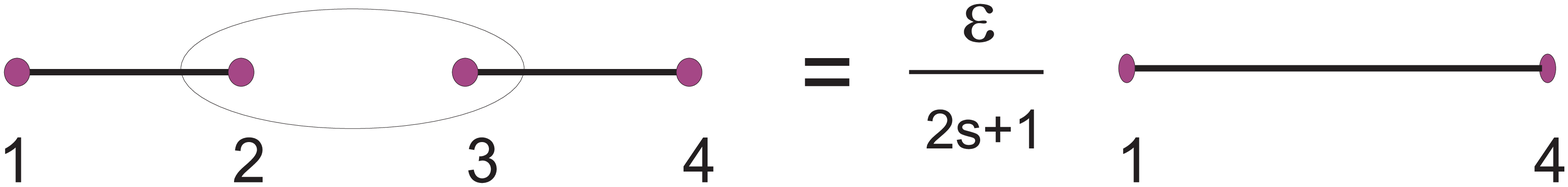}
   \caption{Color Online. The graphical representation of equation (\ref{SingletProduct}).
   A bulb around 2 and 3, means that the state
   has been multiplied from the left by $_{2,3}\la S|$, leaving the singlet on the right hand side. Note that the labels
   1, 2, 3 and 4, denote any four sites.}
   \label{SP}
\end{figure}

\begin{figure}[t]
 \centering
   \includegraphics[width=13cm,height=1cm,angle=0]{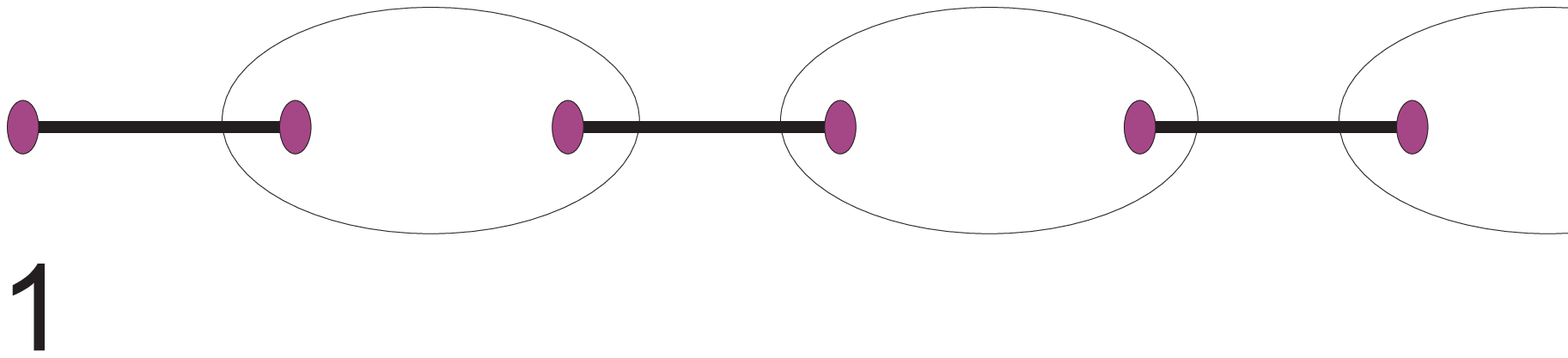}
   \caption{Color Online. The repeated use of the graph (\ref{SP}) for calculating the cross product terms $\la \phi_1|\phi_2\ra$.}
   \label{RSP}
\end{figure}

This relation which we will use repeatedly in the following is
depicted graphically in figure (\ref{SP}). Here a bulb around two
sites $2,3$ means that it has been multiplied from the left by a
singlet $_{_{2,3}}\la S|$. Repeatedly using equation
(\ref{SingletProduct}) or the graph (\ref{SP}), as in figure
(\ref{RSP}), will give \be
\la\phi_1|\phi_2\ra=\frac{\epsilon^N}{(2s+1)^{N-1}} \ee from which
we obtain the normalization \be\label{6}
\la\Psi^{\pm}|\Psi^{\pm}\ra=2(1\pm\frac{\epsilon^N}{(2s+1)^{N-1}}).
\ee In order to find the correlations we use the following
equations, \be _{_{12}}\la S|{\bf s}_1\cdot {\bf
s}_2|S\ra_{_{12}}=-s(s+1),\ee and \be _{_{23}}\la S|{\bf s}_1\cdot
{\bf s}_2|S\ra_{_{12}}|S\ra_{34}=-s(s+1) \ _{_{23}}\la
S|S\ra_{_{12}}|S\ra_{_{34}}=-\frac{s(s+1)\epsilon^N}{2s+1}|S\ra_{_{14}},
\ee which readily gives
\begin{equation}\label{1}
    \la \phi_1|{\bf s}_1\cdot {\bf s}_2|\phi_1\ra=-s(s+1),\h \la \phi_2|{\bf s}_1\cdot {\bf
    s}_2|\phi_2\ra=0.
\end{equation}
Again the cross-product terms is calculated with the help of graph
(\ref{RSP}),
\begin{equation}\label{1}
    \la \phi_1|{\bf s}_1\cdot {\bf s}_2|\phi_2\ra=-s(s+1)\la
    \phi_1|\phi_2\ra =-s(s+1)\frac{\epsilon^N}{(2s+1)^{N-1}}.
\end{equation}
Putting these together we find the final form of the correlation
functions:
\begin{equation}\label{psipsi}
    \frac{\la \Psi^{\pm}|{\bf s}_1\cdot {\bf s}_2|\Psi^{\pm}\ra}{\la \Psi^{\pm}|\Psi^{\pm}\ra} =
    -\frac{s(s+1)}{2}\frac{(2s+1)^{N-1}\pm 2\epsilon^N}{(2s+1)^{N-1}\pm \epsilon^N}.
\end{equation}

To construct the parent Hamiltonian of such states, we use
(\ref{cAA}) and find the null-space of the reduced density matrices
of three consecutive sites $\Delta_3$, (two consecutive sites have
no non-trivial null space in this model). From (\ref{defA}) we have
\begin{equation}\label{AAA}
    A_{m}A_{m'}A_{m''}=(-1)^{s-m'}\delta_{m',-m''}|m\ra\la
    \tilde{0}|+\delta_{m',-m}(-1)^{2s-m-m''}|\tilde{0}\ra\la -m''|.
\end{equation}
To find the null space $\Delta_3,$ we need to solve the matrix
equation
\begin{equation}\label{ccc}
    \sum_{m,m',m''}c_{m,m',m''}A_mA_{m'}A_{m''}=0,
\end{equation}
which yields the following conditions:
\begin{eqnarray}\label{ccc} \sum_{m}(-1)^{m} C_{m',m,-m}&=&0 \h \forall m'\cr \sum_m(-1)^m
C_{m,-m,m'}&=&0 \h \forall m'. \end{eqnarray}

These conditions can be re-expressed in a more useful form, namely
the null space $\Delta_3$ is spanned by vectors of the form
\begin{equation}\label{chi}
|\chi\ra=\sum_{m,m',m''}c_{m,m',m''}|m,m',m''\ra
\end{equation}
which are perpendicular to the state (\ref{SSS}) i.e.
\begin{equation}\label{conditions}
    _{12}\la S|\chi\ra=0,\ \ \ \ \ \ _{23}\la S|\chi\ra=0,
\end{equation}
where the subscripts indicate the embedding of $\la S|$ into the
local spaces of three consecutive spins. Note that the factor
$\sqrt{2s+1}$ has been inserted so that the state $|S\ra$ be
normalized.  We will later use these equations to clarify the form
of the Hamiltonian, but first let us derive an explicit form for the
ground
state. \\

One is tempted to ask if $|\phi_1\ra$ or $|\phi_2\ra$ are ground
states separately. The answer is positive. To see this, note that
the Hamiltonian is written in the form
\begin{equation}\label{HH}
    H=\sum_{k=1}^{2N} h_{k,k+1,k+2},
\end{equation}
where $h$ is the sum of projectors on the null space $\Delta_3$,
i.e.
\begin{equation}\label{proj}
    h=\sum_{\a}\lambda_{\a}|e_{\a}\ra\la e_{\a}|, \h |e_{\a}\ra\in \Delta_3.
\end{equation}

Here $\{|e_{\a}\ra\}$ is a basis for $\Delta_{3}$ and from
(\ref{conditions}) we know that $\la e_{\a}|S\ra_{12}=\la
e_{\a}|S\ra_{23}\ \ \ \ \forall \ \ \a$. This implies that
$H|\phi_1\ra=H|\phi_2\ra=0$. Each of the dimerized states
$|\phi_1\ra$ and $|\phi_2\ra$, break the translational symmetry of
the Hamiltonian. Finally let us also derive the parent Hamiltonians
for the simplest cases, namely spin 1/2 which is well-known and
spin-1 Majumdar-Ghosh states.\\

\textbf{a: The parent Hamiltonian for Spin-1/2 dimerized state
}\label{examples}\\

Using the standard notation $|+\ra:=|\frac{1}{2},\frac{1}{2}\ra, \ \
|-\ra:=|\frac{1}{2},\frac{-1}{2}\ra$ we order the states of
auxiliary space, as $\{|+\ra, |\tilde{0}\ra, |-\ra\}$. Then from
(\ref{defA}) we have
\begin{eqnarray}\label{MGmatrices}
 A_+ &=& |+\ra \la \tilde{0}|+|\tilde{0}\ra\la -|=\left(\begin{array}{ccc} 0 & 1 & 0 \\ 0 & 0 & 1 \\ 0 & 0 & 0\end{array}\right)\cr
 A_- &=& |-\ra\la \tilde{0}|-|\tilde{0}\ra\la +|=\left(\begin{array}{ccc} 0 & 0 & 0 \\ -1 & 0 & 0 \\ 0 & 1
 & 0
 \end{array}\right),
\end{eqnarray}
which transforms as a rank $1/2$ tensor with the generators given by
\begin{equation}\label{LGM}
  L_z = \left(\begin{array}{ccc} \frac{1}{2}  & 0 & 0 \\ 0 & 0 & 0 \\ 0 & 0 &  \frac{-1}{2}\end{array}\right),
  \ \ \ \
  L_+ = \left(\begin{array}{ccc} 0 & 0 & 1 \\ 0 & 0 & 0 \\ 0 & 0 & 0
 \end{array}\right),\ \ \ \ \ L_- = \left(\begin{array}{ccc} 0 & 0 & 0 \\ 0 & 0 & 0 \\ 1& 0 & 0
 \end{array}\right).
\end{equation}
The singlet states are $|S\ra=\frac{1}{\sqrt{2}}(|+,-\ra-|-,+\ra)$.
To find the parent Hamiltonian, we should solve equation
(\ref{ccc}), or what is the same thing, find states $|\chi\ra \in
C^2\otimes C^2\otimes C^2$ such that $_{12}\la S|\chi\ra = _{23}\la
S|\chi\ra=0$. It is readily found that there are four such states:

\begin{eqnarray}
  |e_1\ra &=& |+,+,+\ra, \cr
  |e_2\ra &=& \frac{1}{\sqrt{3}}(|+,+,-\ra+|+,-,+\ra+|-,+,+\ra), \cr
  |e_3\ra &=& \frac{1}{\sqrt{3}}(|-,-,+\ra+|-,+,-\ra+|+,-,-\ra), \cr
  |e_4\ra &=& |-,-,-\ra.
\end{eqnarray}
The vectors $|e_i\ra$ form the spin $\frac{3}{2}$ multiplet, and if
they come with the same coefficients in $h$ in (\ref{proj}), the
resulting Hamiltonian will be a scalar. It is known
\cite{AKS3,mpsgeneral} that in this case the parent Hamiltonian will
be the Majumdar-Ghosh Hamiltonian, namely
$$
    H=\sum_{i}2\sigma_i\cdot \sigma_{i+1}+\sigma_i\cdot
    \sigma_{i+2}.
$$

\textbf{b: The parent Hamiltonian for Spin-1 fully dimerized
state}\\

Using the abbreviated notation $|1\ra:=|1,1\ra, \ |0\ra:=|1,0\ra, \
|\overline{1}\ra:=|1,-1\ra$, we have as the singlet state in
$|S\ra\in C^3\otimes C^3$,

\begin{equation}\label{S1}
    |S\ra=\frac{1}{\sqrt{3}}(|1,\overline{1}\ra-|0,0\ra+|\overline{1},1\ra).
\end{equation}
In order to find the null space $\Delta_3$, we note that due to
$su(2)$ symmetry, equation (\ref{symmH}), the basis vectors of
$\Delta_3$ can be grouped into multiplets which transform
irreducibly under $su(2)$. These multiplets come from the
decomposition of $1\otimes 1\otimes 1$ representation, which
decomposes as
\begin{equation}\label{decom}
    1\otimes 1\otimes 1=3\oplus 2\oplus 2'\oplus 1\oplus
1'\oplus 1''\oplus 0.
\end{equation}
However not all the above multiplets belong to $\Delta_3$. In order
to determine those which are, we should check the conditions
(\ref{ccc}). It is sufficient to check these conditions only for the
top state of each multiplet, since symmetry guarantees that the
other states are present in $\Delta_3$, once the top state is
present. With this insight we readily find the multiplets with the
following top states are present in $\Delta_3$:

\begin{eqnarray}
  |e_1\ra &:=& |t_3\ra=|1,1,1\ra \cr
  |e_2\ra &:=& |t_2\ra = |1,1,0\ra-|0,1,1\ra \cr
  |e_3\ra &:=& |t_{2'}\ra = |1,1,0\ra-2|1,0,1\ra+|0,1,1\ra \cr
  |e_4\ra &:=&
  |t_1\ra=|1,0,0\ra+|0,0,1\ra+3|\overline{1},1,1\ra+3|1,1,\overline{1}\ra-2|1,\overline{1},1\ra-4|0,1,0\ra\cr
|e_5\ra&:=&|t_0\ra=
|1,\overline{1},0\ra-|\overline{1},1,0\ra-|1,0,\overline{1}\ra+|\overline{1},0,1\ra+|0,1,\overline{1}\ra-|0,\overline{1},1\ra,
\end{eqnarray}
where $|t_j\ra$ denotes the top state of the spin-$j$
representation. One can verify that these are actually the top
states by checking the equations $L_z|t_j\ra=j|t_j\ra,$  and $  \ \
L_+|t_j\ra=0$ and also that they really belong to $\Delta_3$ by
checking $_{12}\la S|e_i\ra=_{23}\la S|e_i\ra=0 $.

Having 5 different multiplets in the null space, means that the
Hamiltonian has 5 different couplings which can be tuned. Of course
one of the couplings can be set to unity by a choice of energy
scale. Let's call the projectors on the representation space $j$ by
$P_j$. Then the local Hamiltonian $h$ will be
\begin{equation}\label{hs1}
    h=\lambda_0P_0 + \lambda_1 P_1 + \lambda_2 P_2 +
    \lambda'_{2}P_{2'}+\lambda_3 P_3.
\end{equation}

\textbf{Remark:} It is important to note that the MPS formalism,
gives a larger family of parent Hamiltonian than the original AKLT
construction. In fact in the AKLT construction, the presence of
projectors $P_3$, $P_2$ and $P_{2'}$ and $P_0$ is automatic. However
the presence of the new projector $P_1$ is the result of the MPS
formalism.\\

The next step, which is not trivial, is to write the projectors in
terms of local spin operators. The point is that on the
decomposition (\ref{decom}) only some of the representations on the
right hand side belong to $\Delta_3$. For those representations
which occur with multiplicity one, we can easily find the expression
of the corresponding projectors in terms of local spin operators.
Let us denote the sum of spin operators on three sites by ${\bf S}$,
i.e
$${\bf S}:={\bf S}_1+{\bf S}_2+{\bf S}_3.$$ The basis states of the
representations on the right hand side of (\ref{decom}) are such
that they block-diagonalize the generators and hence the operator
${\bf S}\cdot {\bf S}$. Let us denote the projectors on the totality
of spin $j$ representations by ${\cal P}_j$, i.e. ${\cal P}_0=P_0, \
\ {\cal P}_1:=P_1+P_{1'}+P_{1''}, \ \ \ {\cal P}_2=P_2+P_{2'}$, and
${\cal P}_3=P_3$. Then we have the following system of equations
$({\bf S}\cdot {\bf S})^k=\sum_{l=0}^3(l(l+1))^k{\cal P}_{l}, \
k=0,1,2,3$, or more explicitly,
\begin{eqnarray}\label{}
I&=&{\cal P}_3+{\cal P}_2+{\cal P}_1+{\cal P}_0\cr {\bf S}\cdot {\bf
S}&=&12{\cal P}_3+6{\cal P}_2+2{\cal P}_1\cr ({\bf S}\cdot {\bf
S})^2&=&144{\cal P}_3+36{\cal P}_2+4{\cal P}_1\cr ({\bf S}\cdot {\bf
S})^3&=&343{\cal P}_3+125{\cal P}_2+27{\cal P}_1.
\end{eqnarray}
Inverting the above equations we find
\def\ss{{\bf S}\cdot{\bf S}}

\begin{eqnarray}
 720 {\cal P}_3 &=& 12\ss-8(\ss)^2+(\ss)^3 \cr
144{\cal P}_2 &=& -24\ss+14(\ss)^2-(\ss)^3 \cr 80{\cal P}_1 &=&
72\ss-18(\ss)^2+(\ss)^3 \cr 144{\cal P}_0 &=& 144
-108\ss+20(\ss)^2-(\ss)^3.
\end{eqnarray}
A positive linear combination of the projectors ${\cal P}_3, {\cal
P}_2, $ and ${\cal P}_0$ gives a three parameter family of
Hamiltonians. The projector ${\cal P}_1$ should be left out from
this combination, since only one of the spin-$1$ representations
belong to the null space $\Delta_3$. In general those
representations which occur with multiplicity one, can always be
expressed in terms of total spin operator ${\bf S}$ on three sites.
However we can construct a more general family of Hamiltonians by
calculating explicitly all the projectors in (\ref{hs1}) in terms of
the most general set of independent three-body spin operators.  A
straightforward calculation gives the final form of the Hamiltonian
(with the abbreviation $S_{12}:={\bf S}_1\cdot {\bf S}_2$):

\begin{eqnarray}\label{P1}\nonumber
    H = \sum_{i=1}^{2N} J_0 &+&J_1\  S_{i,i+1} + J_2\  S_{i,i+2} + J_3\  S_{i,i+1}^2 + J_4\  S_{i,i+2}^2
    + J_5\  \{S_{i,i+1}, S_{i+1,i+2}\} \cr &&\cr &+& J_6\{  S_{i,i+2}, \{S_{i,i+1},
    S_{i+1,i+2}\}\}+J_7\ \left(S_{i,i+1}S_{i,i+2}S_{i+1,i+2}+S_{i+1,i+2}S_{i,i+2}S_{i,i+1}\right),
\end{eqnarray}
where
\begin{eqnarray}
J_0 &=& -2\lambda_0 + \frac{3}{5} \lambda_1 +
\frac{1}{3}(\lambda_2+\lambda_{2'})+\frac{1}{15}\lambda_3,\cr
 J_1 &=& 2\lambda_0 - \frac{2}{5} \lambda_1 - \frac{1}{3}(\lambda_2+\lambda_{2'})+\frac{11}{15}\lambda_3,\cr
  J_2 &=& -3\lambda_0 + \frac{1}{20} \lambda_1 + \frac{1}{2}(\lambda_2+\lambda_{2'})-\frac{3}{10}\lambda_3,\cr
   J_3 &=& 2\lambda_0 - \frac{13}{20} \lambda_1 - \frac{1}{6}(\lambda_2-\lambda_{2'})+\frac{1}{15}\lambda_3,\cr
    J_4 &=& \lambda_0 + \frac{1}{20} \lambda_1 + \frac{1}{6}(\lambda_2-\lambda_{2'})+\frac{1}{30}\lambda_3,\cr
     J_5 &=& \lambda_0 - \frac{1}{6}(\lambda_2+\lambda_{2'})+\frac{1}{6}\lambda_3,\cr
      J_6 &=& -\lambda_0 + \frac{1}{10} \lambda_1 + \frac{1}{6}\lambda_{2'}-\frac{1}{10}\lambda_3,\cr
        J_7 &=& \lambda_0 - \frac{1}{40} \lambda_1 - \frac{1}{12}(\lambda_2+5\lambda_{2'})+\frac{7}{30}\lambda_3.
\end{eqnarray}

This Hamiltonian may seem complicated and not so interesting from
the physical point of view. However we should note that it has
effectively four adjustable parameters, (sine we can take
$\lambda_0=1$) and by tuning these parameters this Hamiltonian may
come close to physically simple and interesting models. For example
 if we take the parameters as follows:

\begin{equation}\label{para}
    \lambda_0:=1,\ \ \ \  \lambda_1=8+\frac{2}{3}\lambda_2,\ \ \ \ \lambda'_2=12+2\lambda_2, \ \ \ \ \ \lambda_3=18+4\lambda_2,\ \ \ \
\end{equation}
then the couplings $J_3,J_4,J_6,$ and $J_7$ all vanish and the
Hamiltonian finds the following simple form, modulo additive and
positive multiplicative constants

\begin{equation}\label{hsimple}
H= \sum_{i=1}^{2N} {\bf S}_i\cdot {\bf S}_{i+1} + \Delta {\bf
S}_i\cdot {\bf S}_{i+2} + K \{{\bf S}_i\cdot {\bf S}_{i+1}\ ,\ {\bf
S}_{i+1}\cdot {\bf S}_{i+2}\},
\end{equation}
where $\Delta:=\frac{\lambda_2-6}{5\lambda_2+24}$ and
$K=\frac{6+{\lambda_2/2}}{5\lambda_2+24}$. By taking $\lambda_2=6$,
we can further set $\Delta=0$, and $K=\frac{1}{6}$ and hence we can
arrive at
\begin{equation}\label{hsimple}
H= \sum_{i=1}^{2N} {\bf S}_i\cdot {\bf S}_{i+1} + \frac{1}{6} \{{\bf
S}_i\cdot {\bf S}_{i+1}\ ,\ {\bf S}_{i+1}\cdot {\bf S}_{i+2}\},
\end{equation}
or by taking $\lambda_2$ very large, we can come arbitrarily close
to the following Hamiltonian:
\begin{equation}\label{hsimple}
H= \sum_{i=1}^{2N} {\bf S}_i\cdot {\bf S}_{i+1} + \frac{1}{5}{\bf
S}_i\cdot {\bf S}_{i+2}+\frac{1}{10} \{{\bf S}_i\cdot {\bf S}_{i+1}\
,\ {\bf S}_{i+1}\cdot {\bf S}_{i+2}\}.
\end{equation}

\subsection{Examples of AKLT/VBS states}
While the MPS representation may not be a necessity when dealing
with fully dimerized states , such representation
is invaluable when dealing with VBS states.\\

\textbf{Spin 3/2 VBS state} \\

As our last examples, we consider the MP representation of the spin
3/2 VBS state of the form shown in figure
(\ref{PartialDimersSimple}), which is obtained from a dimerized
state with $s=1$ and $s'=1/2$. From (\ref{Vss}) we see that the MP
representation of such a chain is given by the following matrices,
where we have abbreviated $|1,m\ra\rightarrow |m\ra, $ and
$|\frac{1}{2},\pm\frac{1}{2}\ra\rightarrow |\pm\ra$
\begin{eqnarray}
V_{_{3/2,3/2}} &=& |1\ra\la -| +|+\ra\la -1| \cr V_{_{3/2,1/2}} &=&
\sqrt{\frac{2}{3}}|0\ra\la -| + \sqrt{\frac{1}{3}}|-\ra\la -1|-
\sqrt{\frac{1}{3}}|1\ra\la +| - \sqrt{\frac{2}{3}}|+\ra\la 0| \cr
V_{_{3/2,-1/2}}&=& -\sqrt{\frac{2}{3}}|0\ra\la +| -
\sqrt{\frac{1}{3}}|+\ra\la 1| + \sqrt{\frac{1}{3}}|-1\ra\la -|+
\sqrt{\frac{2}{3}}|-\ra\la 0|  \cr V_{_{3/2,-3/2}}&=& -|-1\ra\la +|
+|-\ra\la 1|.
\end{eqnarray}

Note that we use equation (\ref{Vss}) to find the highest-weight
component of this tensor and the rest of the components are derived
by action of $L_-$. In a basis with the order
$|1,1\ra,|1,0\ra,|1,-1\ra, |+\ra,|-\ra $, the 5 dimensional vectors
$V_{3/2,m}$ have the following explicit form:

\begin{equation}\label{V32}
V_{_{3/2,3/2}}=\left(\begin{array}{ccccc} .& .& .& .& 1\\ .& .& .&
.&.
\\ .&. & .& .& .\\ .&. & 1 &. &. \\ .&. & .& .&
.\end{array}\right)\ \ ,\ \ \
V_{_{3/2,1/2}}=\frac{1}{\sqrt{3}}\left(\begin{array}{ccccc} .& .& .&
-1& .\\ .& .& .& .& \sqrt{2}
\\ .&. & .& .& .\\ .&-\sqrt{2} & . &. &. \\ .&. & 1& .&
.\end{array}\right)
\end{equation}

\begin{equation}\label{V32}
V_{_{3/2,-1/2}}=\frac{1}{\sqrt{3}}\left(\begin{array}{ccccc} .& .& .& .& .\\
.& .& .& -\sqrt{2}&.
\\ .&. & .& .& 1\\ 1&. & . &. &. \\ .&-\sqrt{2} & .& .&
.\end{array}\right)\ \ ,\ \ \
V_{_{3/2,-3/2}}=\left(\begin{array}{ccccc} .& .& .& .& .\\ .& .& .&
.& .
\\ .&.& .& -1& .\\ .&. & . &. &. \\ 1&. & .& .&
.\end{array}\right)
\end{equation}

From figure (\ref{PartialDimersSimple}) and the discussion following
it, we see that the translation-invariant parent Hamiltonian
annihilating this state, should be constructed from the projector
$P_3$ onto spin-3 states. In the matrix product formalism, this
means that the null-space of the two-site density matrix, should
contain the multiplet of states which transform as spin 3
representation. In view of (\ref{fusion}), this means that in the
decomposition of quadratic product of tensors $V_{3/2,m}$, into
irreducible representations of $su(2)$, the representation of spin-3
should not appear, i.e. the components of this tensor should
identically vanish. This is indeed the case as one can see from
(\ref{V32}) that $V_{_{3/2,3/2}}^2$ which is the component with
highest weight of spin-3 representation vanishes. The other
components vanish by symmetry.

Equation (\ref{fusion}) generalizes this to arbitrary spins in a
nice way which is exactly what we see in the valence bond picture of
(\ref{OurPartialDimers}). Even more than that, it gives in one shot,
the Hamiltonian which is common to both $h_1$ and $h_2$ in figure
(\ref{OurPartialDimers}).\\

\textbf{The fusion rule of the tensors $V_{s}$}\\

As stated above the properties of valence bonds in the AKLT
formalism are nicely captured in the fusion rule of the tensors
$V_s$, equation (\ref{fusion}). Although the complete proof of
(\ref{fusion}) is possible, we think it is not so illuminating.
Instead we try to illustrate the idea by two simple example.
Consider figure (\ref{OurPartialDimers}), with $s=3/2$ and $s'=1$.
In our picture the VBS state obtained from projection is an MPS with
auxiliary matrices given by $V_{5/2}:=V_{3/2+1}$. Note that we use
$V_{s}$ to denote the totality of matrices $V_{s,m}$, for all $m$.
We want to show that
\begin{equation}\label{V52}
    V_{5/2}\otimes V_{5/2}=V_0\oplus V_1\oplus V_2\oplus V_3,
\end{equation}
that is, we want to show that in the decomposition of the left hand
side the tensors $V_4, $ and $V_5$ do not appear, hence local
Hamiltonian annihilating the state of the lower chain in figure
(\ref{OurPartialDimers}) can be constructed from projectors $ P_4,$
and $P_5$. To prove this we need only show that the highest weight
components of the tensors $V_5,$ and $ V_4, $ in the decomposition
of the left hand side of (\ref{V52}) vanish identically. To this
end, let us write the explicit form of the components of $V_{5/2}$,
obtained from the top component $V_{5/2,5/2}$ by using (\ref{Vss})
and applying the commutation relation $V_{s,m}\propto
[L_-,V_{s,m+1}]$. Ignoring the numerical coefficients and signs in
front of all states on both sides, which are irrelevant for the
following proof, and using the shortened notation
$|s,m\ra\rightarrow |m\ra$ (i.e. $|1/2\ra\equiv |3/2,1/2\ra,
|-1\ra\equiv |1,-1\ra)$, we have
\begin{eqnarray}\nonumber\label{V52}
  V_{5/2,5/2} &=& | \frac{3}{2}\ra\la -1|+
  |1\ra\la \frac{-3}{2}|,\cr &&\cr
V_{5/2,3/2} &=& |\frac{1}{2}\ra\la -1|+|\frac{3}{2}\ra\la 0|+
  |0\ra\la \frac{-3}{2}|+|1\ra\la \frac{-1}{2}|.
\end{eqnarray}

It is now easily seen that $ V_{5/2,5/2}^2=0$, implying that the
highest weight of the $V_5$ vanish. Moreover we see that
$V_{5/2,5/2}V_{5/2,3/2}=V_{5/2,3/2}V_{5/2,5/2}=0$, implying the
highest weight of $V_4$ also vanishes.  This example corresponds to
figure (\ref{OurPartialDimers}) with $s=3/2$ and $s'=1$ (or with 3
and 2 valence bonds in the AKLT construction). There is a very
interesting point here which we should mention. The point is that a
spin 5/2 VBS state can also be constructed in the same way as in
figure (\ref{OurPartialDimers}) with $s=2$ and $s'=1/2$ or as in the
original picture, from partially dimerized states with different
numbers, namely with 4 and 1 valence bonds. Here we expect that the
local Hamiltonian which is used in the construction of
translation-invariant state be constructed only from the projector
$P_5$. This is nicely captured in the fusion rule of our tensors,
$W_{5/2}:=V_{2+1/2}$, which is

\begin{equation}\label{W}
    W_{5/2}\otimes W_{5/2}=W_4\oplus W_3\oplus W_2\oplus W_1\oplus
    W_0.
\end{equation}
In fact we have (again ignoring numerical coefficients on both
sides), and with the same type of shortened notation as in the
previous example,
\begin{eqnarray}\nonumber
  W_{5/2,5/2} &=& |2\ra\la -1/2|+
  |1/2\ra\la -2|,\cr &&\cr
W_{5/2,3/2} &=& |1\ra\la -1/2|+|2,\ra\la 1/2|+
  |-1/2\ra\la -2|+|1/2\ra\la -1|.
\end{eqnarray}

It is now seen that while the top state of $W_5$ is zero, the top
state of $W_4$, that is
$W_{5/2,5/2}W_{5/2,3/2}-W_{5/2,3/2}W_{5/2,5/2}$ is non-vanishing,
proving the fusion rule (\ref{fusion}). This argument can be
generalized to the arbitrary spins $s$ and $s'$, although the proof
will not be more illuminating than the example given above.

\section{Conclusion}\label{conclusion}
The main emphasis of this paper has been on the rotational symmetry
properties of matrix product states. To this end we have constructed
a simple representation of spherical tensors of arbitrary integer or
half integer rank. A spherical tensor of rank $s$ is represented in
a $2s+2$ dimensional space, hence the dimension of space, increases
only linearly with the rank of the tensor. The introduction of these
tensors have made possible a unified approach toward fully
dimerized, and partially dimerized or AKLT/VBS states. In this way
we have been able to find a matrix product representation for all
the variety of valence bond states introduced in the original AKLT
paper.  Having such a matrix product representation makes the
calculation of many properties of such states, specially the
non-dimerized states quite easy and straightforward.  Moreover a MPS
representation is more powerful, since it will give a larger family
of Hamiltonians compared with the AKLT construction, since it allows
to include more projectors in the local Hamiltonian. This will then
lead to more flexibility in approximating realistic interactions
with parent Hamiltonians of matrix product states. We have
demonstrated this for a spin-1 family of Hamiltonians with nearest
and next-nearest neighbor interactions. Finally we should remind
that the above constructions can be generalized to other symmetry
groups like $su(n)$. At least for a self-conjugate representation of
$su(n)$, whose weight diagram is symmetric under reflection, then we
can define tensor operators in exactly the same way as in equation
(\ref{defA}), namely:

\begin{equation}\label{m}
    A_{{\bf m}}:=|{\bf m}\ra \la \tilde{0}|+|\tilde{0}\ra \la -{\bf m}|,
\end{equation}
whee $|{\bf m}\ra$ is the $n-1$ dimensional weight vector of that
representation. Such a MPS representation may be useful for example
in recent considerations of AKLT models as in \cite{Korepin,
Stephen} where $su(2)$ valence bonds have been replaced with su(n)
valence bonds, or in \cite{Stephen}, where trimmer ground states
with $su(3)$ symmetry have been studied.

\section{Acknowledgements} We would like to thank Kh. Heshami for a
very valuable discussion on the AKLT construction, and I. P.
McCulloch for instructive email correspondences on spherical
tensors.

{}


\begin{thebibliography}{}

\bibitem{Baxter} R. J. Baxter, {\it{Exactly Solved Models in Statistical
Mechanics}}, Academic Press, London, 1982.

\bibitem{LiebMattis} E. H. Lieb, and D. C. Mattis (Eds.) {\it{Mathematical Physics in One
Dimension}}, Academic Press, N.Y. 1966.

\bibitem{Mattis} E. Lieb, T. Schultz, and D. Mattis, Annals of
Physics, {\bf 16} 407-466 (1966).

\bibitem{Bethe} H. A. Bethe, {\it{Proc. Roy. Soc.}} {\bf A 150}, 522
(1935).

\bibitem{AKLT}I. Affleck, T. Kennedy, E.H. Lieb, H. Tasaki, Commun.Math. Phys. \textbf{115}, 477 (1988);
I. Affleck, E.H. Lieb, T. Kennedy, H. Tasaki, Phys. Rev. Lett.
\textbf{59}, 799 (1987).

\bibitem{Majumdar}C. K. Majumdar, J. Phys. C 3. 911(1969);
C. K. Majumdar and D. P. Ghosh, J. Math. Phys. 10 (1969)1388; C. K.
Majumdar and D. P. Ghosh, J. Math. Phys. 10 (1969)1399.

\bibitem{korepin1}  H. Fan, V. E. Korepin, and V. Roychowdhury, Physical Review Letts., {\bf 93},
22, 227203, (2004);   H. Fan, et al. Phys. Rev. {\bf B 76}, 014428
(2007).

\bibitem{akm} S. Alipour, V. Karimipour and L. Memarzadeh, Phys. Rev. A \textbf{75}, 052322
(2007).

\bibitem{Osterloh} A. Osterloh, L. Amico, G. Falci, and R. Fazio,
    Nature 416, 608 (2002).

\bibitem{NielsenOsborne} T. Osborne, M. Nielsen, Phys. Rev. A, {\bf 66}, 032110 (2002).

\bibitem{zit2d} M. A. Ahrens, A. Schadschneider, and J. Zittartz, "Exact ground states of quantum spin-2 models on the
hexagonal lattice" e-print arXiv:cond-mat/0504023.

\bibitem{peps}  D. Perez-Garcia, F. Verstraete, J. I. Cirac, and M. M. Wolf
"PEPS as unique ground states of local Hamiltonians", e-print
arXiv:0707.2260.

\bibitem{FCS1} M. Fannes, B. Nachtergaele, R.F. Werner, Commun. Math.
                 Phys. \textbf{144}, 443 (1992).

\bibitem{FCS2} A. Kl\"{u}mper, A. Schadschneider, and J. Zittartz, J. Phys. A \textbf{24},
L955 (1991); Z. Phys. B \textbf{87}, 281 (1992).

\bibitem{zitt1}A. Kl\"{u}mper, A. Schadschneider, and J. Zittartz, Europhys. Lett. \textbf{24}, 293 (1993).

\bibitem{MPS1}H. Niggemann, J. Zittartz, J. Phys. A: Math. Gen.
\textbf{31}, p. 9819-9828 (1998); H. Niggemann, A. Kl\"{u}mper, J.
Zittartz, Z. Phys. B \textbf{104}, 103
             (1997).

\bibitem{MPS2} M. A. Ahrens, A. Schadschneider, and J. Zittartz,
Europhys. Lett. {\bf 59} 6, 889 (2002); E. Bartel, A. Schadschneider
and J. Zittartz, Eur. Phys. Jour. B, {\bf 31}, 2, 209-216 (2003).

\bibitem{MPS3} M. M. Wolf, G. Ortiz, F. Verstraete and I.
Cirac, Phys. Rev. Lett.{\bf 97}, 110403 (2006).

\bibitem{MPS4}A. K. Kolezhuk and H. J. Mikeska, Phys. Rev.
Lett. {\bf 80}, 2709 (1998); Int. J. Mod. Phys. B, Vol.12, 2325-2348
(1998).

\bibitem{Anders}S. Anders et al. Phys. Rev. Lett. {\bf 97}, 107206
(2006); F. Verstraete and J. I. Cirac, Phys. Rev. B {\bf 73}, 094423
(2006).

\bibitem{AKS1}M. Asoudeh, V. Karimipour, and A. Sadrolashrafi, Phys. Rev. B \textbf{75}, 224427
(2007); Phys. Rev. A \textbf{76}, 012320(2007).

\bibitem{AKS3} M. Asoudeh, V. Karimipour, and A. Sadrolashrafi, Phys. Rev. B 76, 064433 (2007).

\bibitem{mpsgeneral} D. Perez-Garcia, F. Verstraete, M.M. Wolf, and J.I.
Cirac, Journal of Quantum Inf. Comput. {\bf 7}, 401 (2007).

\bibitem{zittartz32} H. Niggemann, and J. Zittartz, Z. Phys. B {\bf 101} (2), p. 289-297
(1996), Freitag, W.-D, and M\"{u}ller-Hartmann, E., Z. Phys. {\bf B
83},381 (1991),E., Z. Phys. {\bf B 88}, 279 (1992).

\bibitem{Del} J.M.Roman, G.Sierra, J.Dukelsky, and  M.A. Martin-Delgado, "
The Matrix Product Approach to Quantum Spin Ladders", e-print
cond-mat/9802150v1.

\bibitem{KolM} Kolezhuk et al. Physical Review {\bf B 55}, 3336(1997).

\bibitem{dmrg} Ostlund and Rommer Phys. Rev. Lett. {\bf 75}, 3537 (1995).

\bibitem{dmrgcirac} F.
Verstraete, D. Porras, and J. I. Cirac, Phys. Rev. Lett. {\bf 93},
227205 (2004).

\bibitem{McGulloch}I. P. McCulloch, J. Stat. Mech. (2007) P10014.


\bibitem{as} K. Totsuka and M. Suzuki, J. Phys. Cond.  Matt. 7, 1639 (1995); J.
Phys. A. 27, 6443 (1994).

\bibitem{shor} D. Nagaj, et al. "The Quantum Transverse Field Ising Model on an
Infinite Tree from Matrix Product States", e-print arXiv:0712.1806.

\bibitem{abk} S. Alipour, S. Baghbanzadeh, and  V. Karimipour, "Exact symmetry breaking ground states for quantum spin
chains", e-print arXiv:0801.1247.

\bibitem{Korepin} H. Katsura, T. Hirano, and V. E. Korepin, "Entanglement in an SU(n) Valence-Bond-Solid State Authors:
", e-print, arXiv:0711.3882.

\bibitem{Stephen}  M. Greiter, and S. Rachel, Phys. Rev. {\bf B 75}, 184441
(2007);  M. Greiter, S. Rachel, and D. Schuricht, Phys. Rev. {\bf B
75}, 060401(R) (2007).



\end{thebibliography}
\end{document}